# COMPOSER2VEC: A CONTINUOUS EMBEDDING SPACE OF COMPOSER STYLE LEARNED FROM SYMBOLIC MELODY GENERATION

*Sakutaro Nishio, Osamu Ichikawa*

Shiga University, Japan

## ABSTRACT

We analyze the composer embeddings learned by a composer-conditioned Transformer as a continuous latent space of compositional style, rather than merely as an internal representation for generation. A model that recursively predicts melody continuations was trained on melodic sequences extracted from MIDI data, conditioned on composer identity (124 composers). Principal component analysis of the learned composer embedding matrix (124×128) shows that the first principal component correlates strongly with composer birth year ($r = -0.884$, $p < 0.001$, $n = 123$), a stronger correlation than we obtain by applying the same PC1–birth-year analysis to existing general-purpose audio-text embeddings (CLAP, MuQ-MuLan) trained on unrelated audio-text corpora, not on symbolic melody generation. A shuffle test (2,000 permutations) confirms that the Silhouette score for stylistic-period labels is statistically significant (0.0110, $p < 0.001$). We further show that vector arithmetic in the embedding space captures meaningful stylistic relationships between composers. These results suggest that composer embeddings, learned without supervision beyond composer identity, form an interpretable latent space that captures musical-historical structure.

## 1. INTRODUCTION

Deep-learning-based automatic composition has attracted growing attention [1]-[3], and many generative models condition on side information via embeddings [4], [5]. However, most prior work on composers formulates the problem as classification [6]-[8] rather than treating stylistic or historical relationships as a continuous structure.

This paper analyzes composer embeddings not as a mere internal representation, but as a continuous latent space of compositional style. Because the embedding is trained only by minimizing generation error, without supervision on era, school, or nationality, any proximity between composers of similar period or style must be a byproduct of training.

We built a composer-conditioned Transformer that recursively generates melody continuations from a melodic sequence and a composer ID, then analyze whether the resulting embeddings capture musical similarity and historical structure without explicit supervision — a framework we call Composer2Vec.

## 2. RELATED WORK

Since Transformers were introduced to symbolic music generation [9], similar architectures have been widely used for generation and representation learning [10]-[12], including embeddings conditioned on composer ID or style [13]-[15]. Encoding performance style with a Transformer autoencoder [16] similarly extracts a latent space, but targets performers rather than composers; composer-classification work [8] remains limited to discrete labels. Composer Vector [17] applies style control post hoc, whereas we jointly learn and analyze the embedding during generative training.

## 3. DATASET AND PREPROCESSING

We used 8,972 of 9,070 MIDI files (124 composers) successfully parsed after cleaning, collected from the archive Kunst der Fuge [18]. Each piece was quantized to a binary piano roll (drum parts excluded) and transposed to a common key (C major/A minor). Using a time resolution of 1/12 of a quarter note, we extracted 8-measure (384-frame) segments with a stride of 192 frames, yielding 277,718 segments, split 8:2 by stratified sampling on composer label. An extra pitch dimension flags rests (set to 1 when no pitch is active in a frame).

The collected MIDI files mixed accurate hand-entered transcriptions with auto-transcribed performances containing excessive ornamental notes unsuitable for training. We built a rule to exclude the latter using two features: the fraction of note onsets aligned to a 1/48-note grid (grid_snap_ratio) and the number of distinct note durations (unique_beats_rounded_count). Using 100 manually labeled songs for validation (Fig. 1), we obtained the linear rule unique_beats_rounded_count $\leq$ 1142.8 $\times$ grid_snap_ratio $-$ 171.4, which achieved higher recall on garbage data than an RBF-kernel SVM (0.90 vs. 0.70) and was adopted; its coefficients were fit visually on the 100-song set and not re-fit on the full collection. This recall was measured on the same 100-song set used to fit the rule; given the limited number of labeled examples (10 garbage songs), it should be read as a rough comparison against the SVM baseline rather than an estimate of held-out generalization performance.

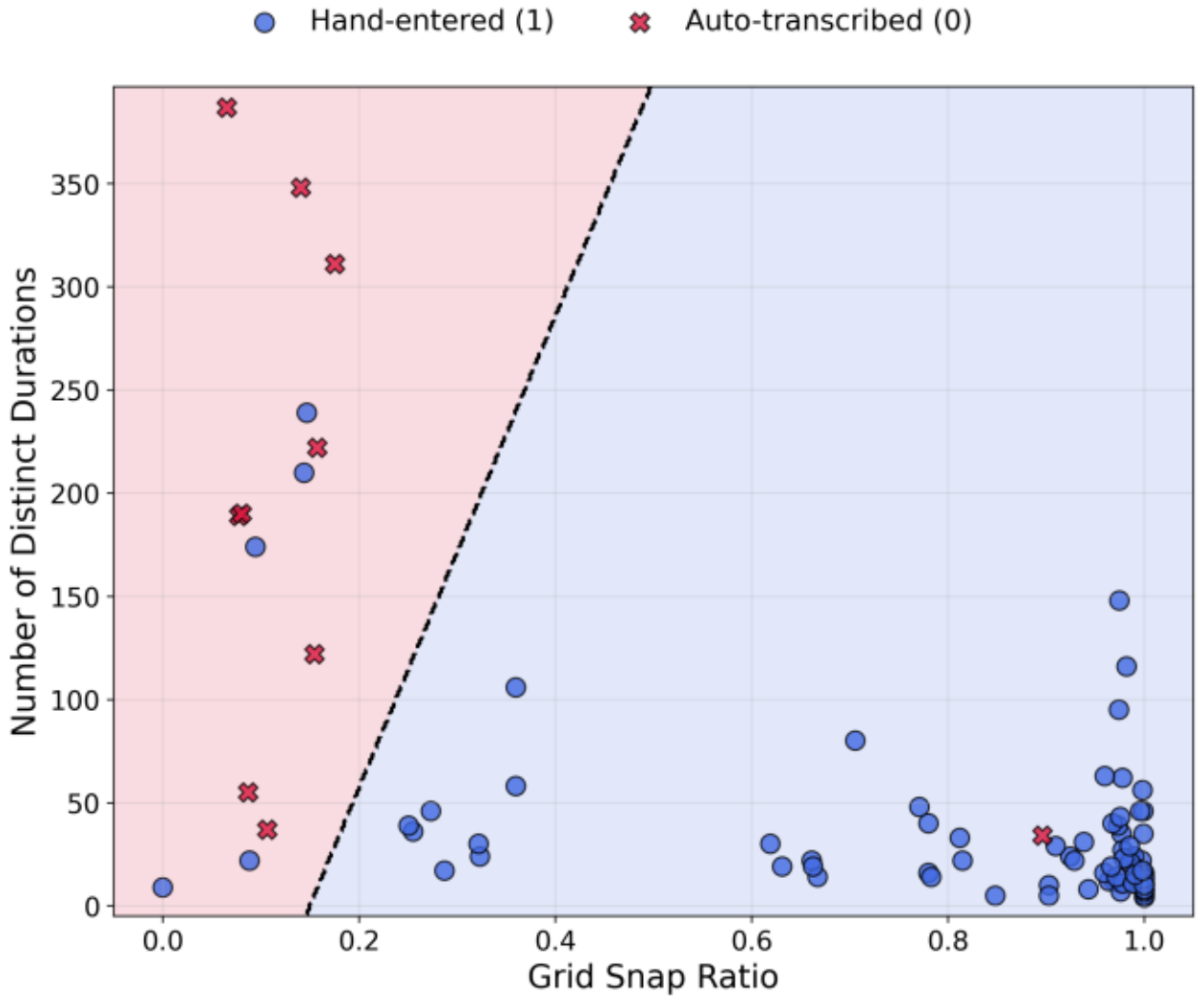

*Fig. 1. Decision boundary and data distribution for the 100 validation songs (x-axis: Grid Snap Ratio; y-axis: Number of Distinct Durations).*

## 4. MODEL

Fig. 2 shows the architecture. The model takes a binarized piano roll as input, with 48 pitch dimensions (four octaves) and 384 time steps at 1/48-note resolution (roughly 8 measures; the finer resolution allows triplets to be represented). Velocity and onset information are not retained, so repeated notes cannot be distinguished from a single sustained note. Each quantized pitch vector is projected to 128 dimensions, to which positional embeddings and a one-hot composer-ID embedding are added at every time step. The model consists of two 128-dimensional Transformer layers with 4-head causal self-attention. The output layer independently predicts, for each time step and pitch, whether a note is active (sigmoid activation, binary cross-entropy loss). At generation time, outputs are stochastically binarized via Bernoulli sampling; if the rest dimension is active for a frame, all pitch dimensions for that frame are set to zero.

Training used causal teacher forcing with AdamW (lr = 1e-3, weight_decay = 1e-2, dropout = 0.2, batch size = 1024), with early stopping (patience = 5, up to 300 epochs). The sequence length was fixed at 386 (bos+192+sep+192); bos/eos/sep are zero-embedded and effectively inactive. In the current implementation a separator token is inserted between the first and second four-measure halves; we plan to remove it in future work. The random seed was fixed; the adopted model is from epoch 300 (train loss 0.0338, test loss 0.0324).

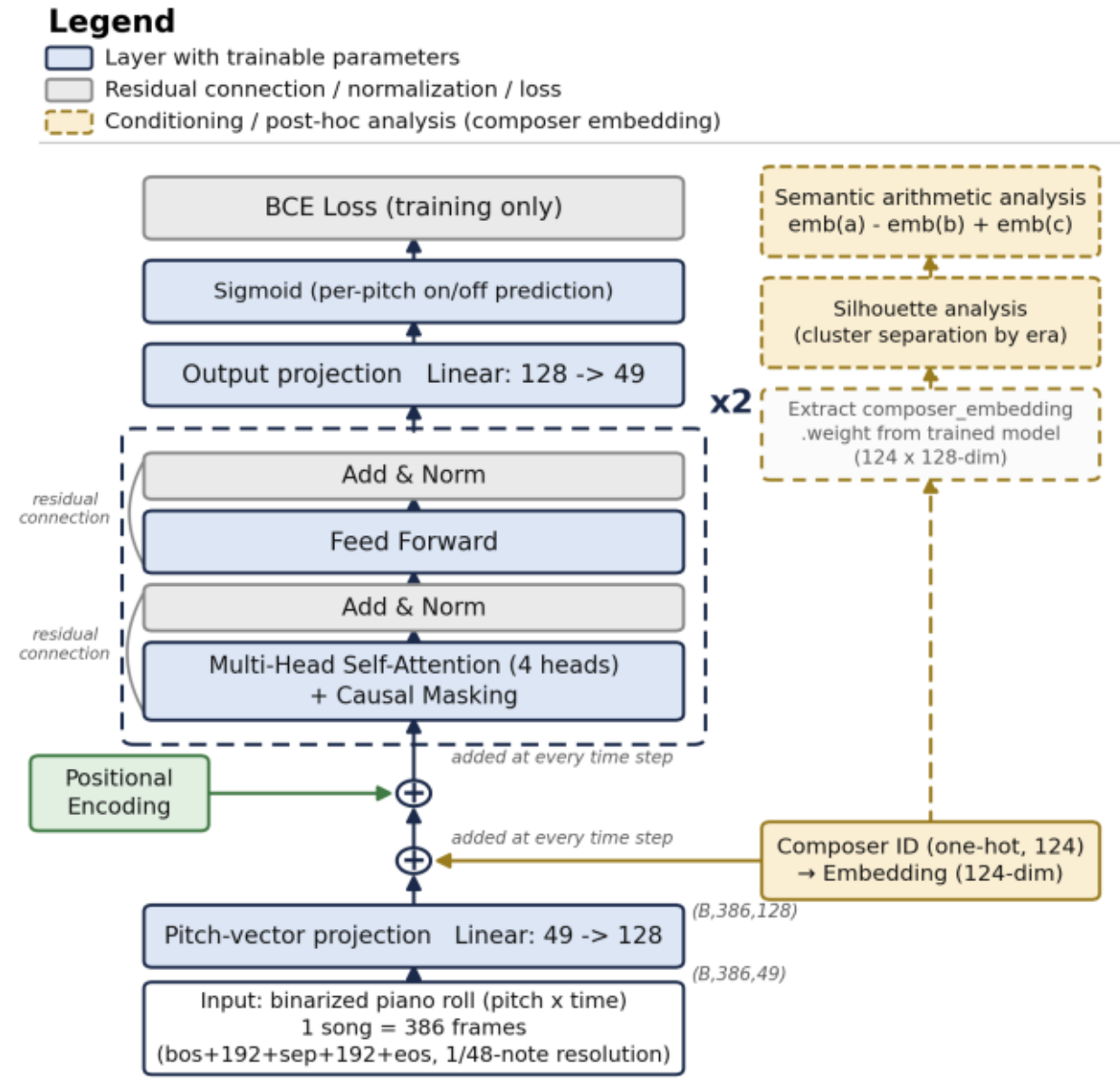

*Fig. 2. Model architecture. Composer-ID embedding and positional encoding are added after the pitch-vector projection, followed by two causally masked self-attention / feed-forward blocks. B denotes batch size.*

## 5. ANALYSIS OF LEARNED COMPOSER EMBEDDINGS

PCA on the learned composer embedding matrix (124×128) shows that the first principal component (PC1) alone explains 21.5% of total variance. For the 123 composers with known birth years, PC1 correlates strongly and negatively with birth year ($r = -0.884$, $p < 0.001$, $n = 123$; Fig. 3), indicating that a single axis tracking each composer's active period emerges as the dominant direction of variation, despite no supervision beyond composer ID.

This temporal axis is also visible when the full 128-D embedding space is projected to 2-D via t-SNE, colored by stylistic period (Fig. 4). The quantitative evaluations below are computed directly in this original 128-D space rather than on the 2-D t-SNE projection. In Fig. 4, Renaissance and Baroque composers occupy the upper region, Romantic and Modern composers the lower region, with Classical composers in between.

We quantified this separation via the Silhouette score on period labels, assessed by a shuffle test (2,000 permutations). The observed score (0.0110) significantly exceeded the shuffled null distribution (mean = -0.0469, SD = 0.0134; Z = 4.33, $p < 0.001$).

To test whether this separation tracks actual historical distance, we correlated, over all 10 period pairs (Baroque, Classical, Romantic, Modern, Renaissance), the mean birth-year gap with the pairwise Silhouette score. We found a strong positive correlation ($r = 0.885$, $p = 0.0007$, $n = 10$; Fig. 5): temporally close pairs (e.g., Romantic-Modern, 52-year gap) showed low separation (0.076), while distant pairs (e.g., Modern-Renaissance, 372-year gap) showed high separation (0.255), consistent with continuous stylistic change across periods. Because each of the 5 periods contributes to 4 of the 10 pairs, these pairs are not statistically independent, so this p-value should be read as a descriptive indication of the trend rather than a rigorously validated significance test.

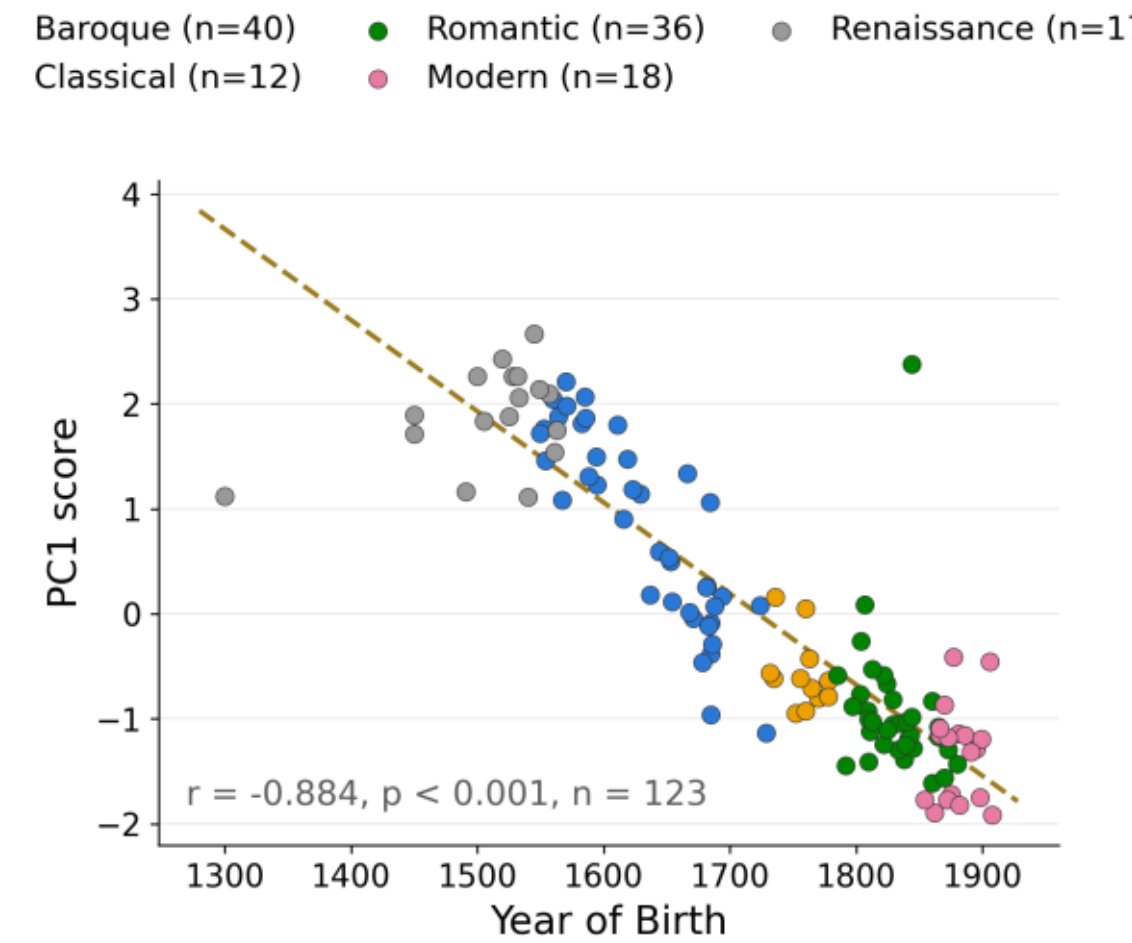

*Fig. 3. First principal component (PC1) of the composer embedding vs. birth year (r = -0.884, p < 0.001, n = 123). Color indicates stylistic period.*

### 5.1. Semantic arithmetic

Following the analogy operations used for word embeddings [19], emb(beethoven) − emb(haydn) + emb(schumann) yields brahms as the nearest neighbor (cosine similarity 0.787), with all top-5 candidates being Romantic composers — consistent with Beethoven's role as a bridge from the Classical to the Romantic era. The independent operation emb(bach-js) − emb(handel) + emb(haydn) ranks beethoven second (0.618), after hummel (0.766), again reflecting this bridging role.

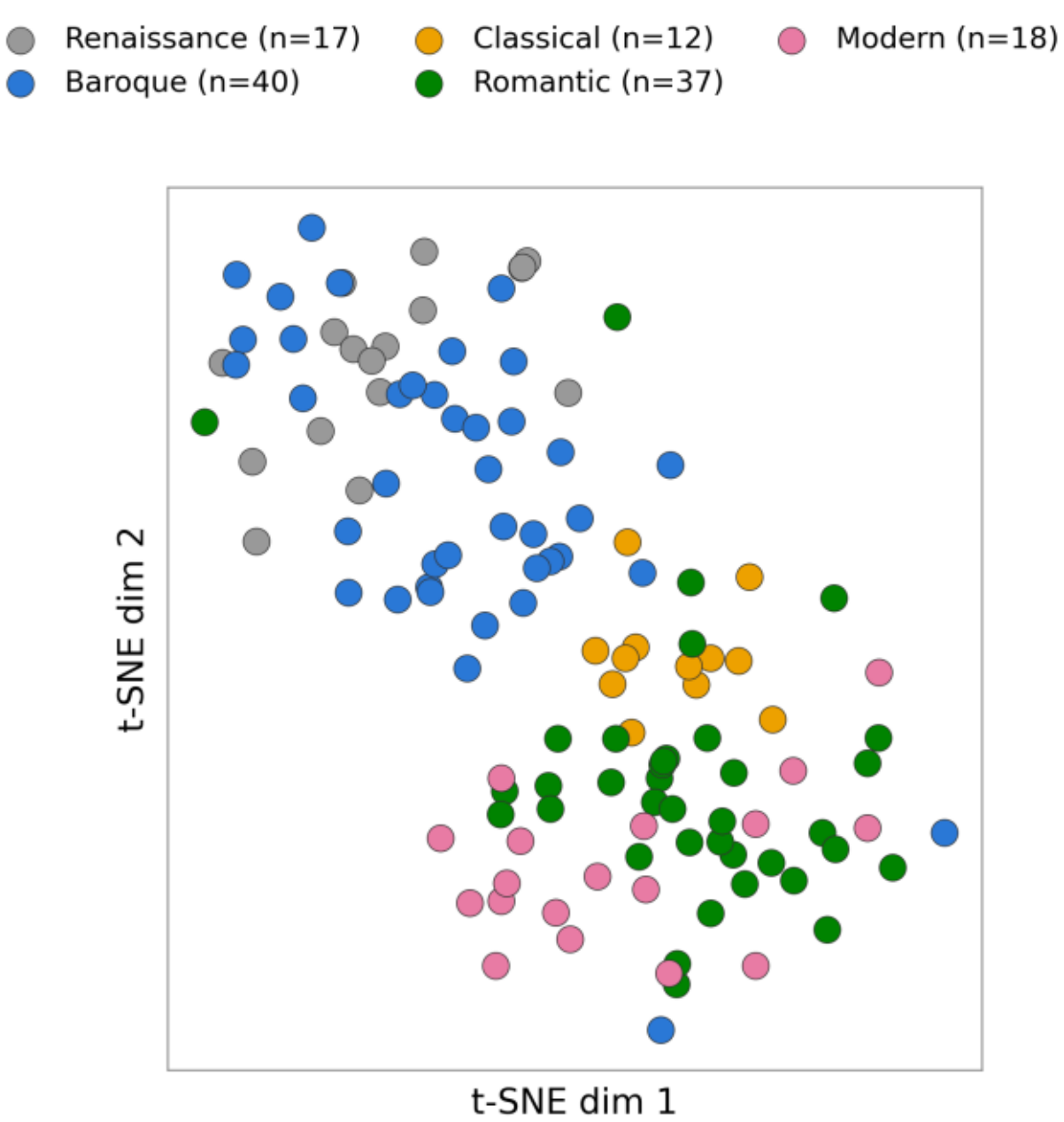

*Fig. 4. t-SNE visualization of the composer embedding (124×128), colored by stylistic period.*

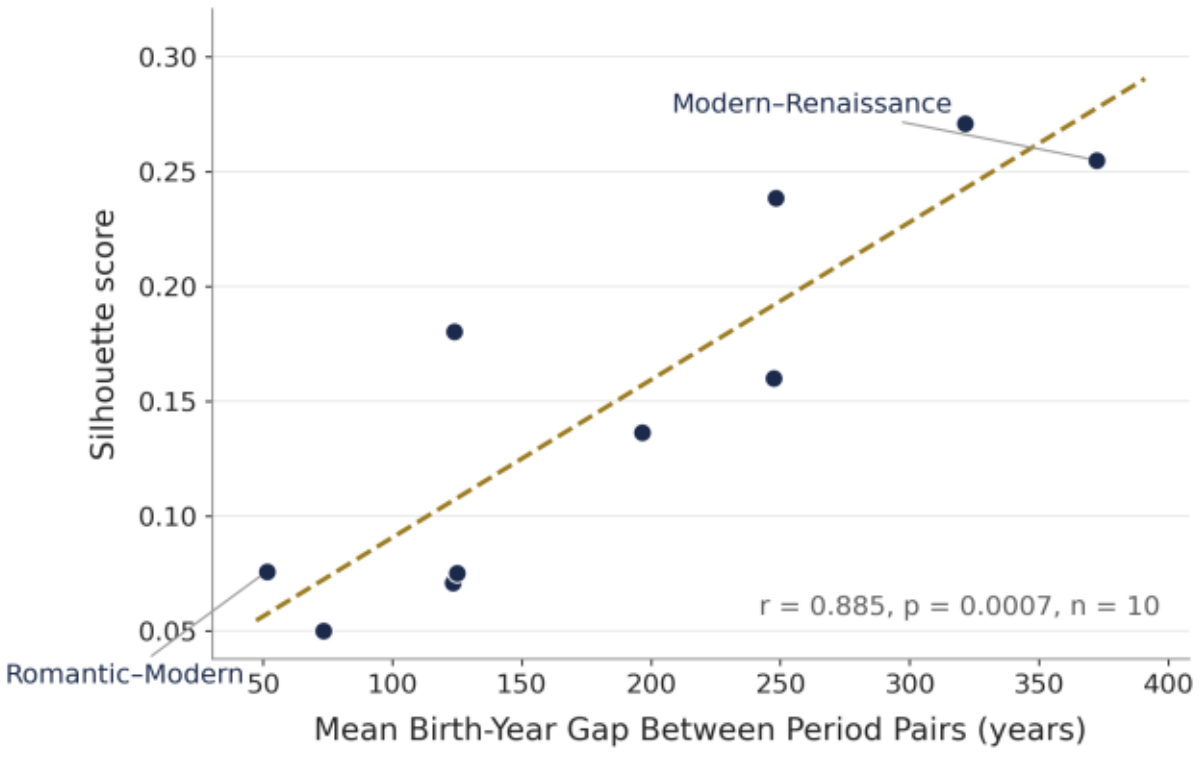

*Fig. 5. Mean birth-year gap vs. Silhouette score for 10 period pairs (r = 0.885, p = 0.0007). Temporally closer pairs show lower separation.*

## 6. COMPARISON WITH CLAP AND MUQ-MULAN

We compared Composer2Vec with two pretrained audio embedding models, CLAP [20] and MuQ-MuLan [21], [22]. Since these models take audio waveforms rather than symbolic MIDI as input, we synthesized audio from the same MIDI collection and recomputed embeddings from the resulting waveforms, evaluating all three models with identical protocols.

Table 1 shows the PC1–birth-year correlation for each model. All three are significant, but Composer2Vec is strongest ($|r| = 0.884$), followed by MuQ-MuLan (0.679) and CLAP (0.532).

Table 1. PC1–birth-year correlation by model ($n = 123$).

| Model | r | p |
|---|---|---|
| Composer2Vec | −0.884 | < 0.001 |
| CLAP | +0.532 | < 0.001 |
| MuQ-MuLan | −0.679 | < 0.001 |

Table 2 shows the period Silhouette score and shuffle-test p-value. Composer2Vec ($p < 0.001$) and MuQ-MuLan ($p < 0.001$) are significant; CLAP is not ($p = 0.548$). Composer2Vec also attains the highest observed score.

Table 2. Period Silhouette score and shuffle-test p-value ($n = 2{,}000$).

| Model | Silhouette | p (shuffle) |
|---|---|---|
| Composer2Vec | 0.0110 | < 0.001 |
| CLAP | −0.1174 | 0.548 |
| MuQ-MuLan | 0.0006 | < 0.001 |

For semantic arithmetic, we applied PCA to each model's embeddings and restricted the arithmetic to the top-10 principal components, ensuring a fair comparison across differing embedding dimensionalities (cumulative explained variance: 70.4% for Composer2Vec, 94.7% for CLAP, 96.0% for MuQ-MuLan). Table 3 shows the top-1 candidate for two representative operations.

Table 3. Top-1 nearest neighbor for two semantic-arithmetic operations (PCA top-10 components; cosine similarity in parentheses).

| Combination | Composer2Vec | CLAP | MuQ-MuLan |
|---|---|---|---|
| beethoven − haydn + schumann | brahms (0.865) | brahms (0.983) | janacek (0.906) |
| wagner − brahms + debussy | satie (0.596) | rubinstein (0.779) | rossini (0.923) |

Agreement varies by operation: for beethoven − haydn + schumann, Composer2Vec and CLAP both rank brahms first, consistent with Beethoven's bridging role, whereas for wagner − brahms + debussy the three models diverge. We suspect this reflects the operations' different character: the former traces a largely chronological, widely agreed-upon transition, while the latter spans a more contested stylistic axis together with strong timbral and orchestrational cues that a purely symbolic, melody-only representation such as ours does not encode; audio-based models may instead be driven by orchestration-related similarity. A more rigorous, music-theoretically grounded interpretation is left to future work with domain experts.

## 7. CONCLUSION

We built a composer-conditioned Transformer for melody continuation and analyzed the resulting composer embeddings. Despite no supervision beyond composer identity, PC1 correlated strongly with birth year ($r = -0.884$, $p < 0.001$), and the period Silhouette score significantly exceeded a shuffled baseline. Embedding arithmetic further revealed a bridging relationship between the Classical and Romantic eras via Beethoven. Future work will further exploit semantic arithmetic and improve the input representation and model architecture.

## 8. ACKNOWLEDGEMENTS

This work was supported by JSPS KAKENHI Grant Number 25K15402.